\documentclass[12pt,preprint]{aastex}
\def\be{\begin{equation}}
\def\ee{\end{equation}}

\newcommand{\msun}{{\rm M}_{\sun}}

\catcode`\@=11 % This allows us to modify PLAIN macros.
\def\@versim#1#2{\vcenter{\offinterlineskip
        \ialign{$\m@th#1\hfil##\hfil$\crcr#2\crcr\sim\crcr } }}

\newcommand{\xte}{{\em RXTE}}

\shorttitle{****}
\shortauthors{****}
\begin{document}
%\date{}

\title{Modeling the hard states of XTE J1550--564 during its 2000 outburst}

\author{Feng Yuan\altaffilmark{1,2,5}, Andrzej A. Zdziarski\altaffilmark{3},
Yongquan Xue\altaffilmark{2}, Xue-Bing Wu\altaffilmark{4}}
\altaffiltext{1}{Shanghai Astronomical Observatory, Chinese Academy of Sciences,
80 Nandan Road, Shanghai 200030, China; fyuan@shao.ac.cn}
\altaffiltext{2}{Department of Physics, Purdue University, West Lafayette,
IN 47907}
\altaffiltext{3}{Centrum Astronomiczne im.\ M. Kopernika, Bartycka 18, 
00-716 Warszawa, Poland}
\altaffiltext{4}{Department of Astronomy, Peking University, Beijing 100871, 
China}
\altaffiltext{5}{Joint Institute for Galaxy and Cosmology (JOINGC) of 
SHAO and USTC}

\begin{abstract}
We study hard states of the black-hole binary XTE J1550--564 during its 2000 outburst. In order to explain those states at their highest luminosities, 
$L\sim 10\%$ of the Eddington luminosity, $L_{\rm E}$, we propose a specific hot 
accretion flow model. We point out that the highest values of the hard-state $L$ 
are substantially above the $L$ an advection-dominated accretion flow (ADAF) can 
produce, $\sim 0.4\alpha^2 L_{\rm E}$, which is only $\sim (3$--$4)\%L_{\rm E}$ 
even for $\alpha$ as high as $0.3$. On the other hand, we successfully explain 
the hard states with $L\sim (4$--$10)\%$ using the luminous hot accretion flow 
(LHAF) model. As $10\%L_{\rm E}$ is also roughly the highest luminosity an 
LHAF can produce, such an agreement between the predicted and observed highest 
luminosities provides by itself strong support for this model. Then, we study 
multi-waveband spectral variability during the 2000 outburst. In addition to 
the primary maxima in the optical light curves, secondary maxima were detected 
after the transition from the very high state to the hard state. We show that 
the secondary maxima are well modeled by synchrotron emission from a jet formed 
during the state transition. We argue that the absence of the corresponding 
secondary peak in the X-ray light curve indicates that the X-ray jet emission, regardless of its radiative process, synchrotron or its Comptonization, is not important in the hard state compared to the emission from the accretion flow.

\end{abstract}


\section{Introduction}
\label{intro}

Black hole X-ray binaries (BHXBs) appear in five main spectral states, namely
quiescent, low/hard, intermediate, high/soft, and very high ones (e.g., McClintock \& Remillard 2006, though they prefer a different nomenclature). The origin of the X-ray spectrum of the hard state presents a highly interesting problem. The presence of a universal high-energy spectral cutoff above $\sim$100--200 keV in the hard state points out to the X-ray emission coming from thermal Comptonization in a hot accretion corona with an electron temperature of $\sim 10^9\,{\rm K}\simeq 10^2\,{\rm keV}/k$ (see, e.g., Zdziarski 2000; Zdziarski \& Gierli\'{n}ski 2004 for reviews). We note that an alternative jet model (Markoff, Nowak \& Wilms 2005), though also relying on thermal Comptonization as the dominant component at photon energies $\ga 10$ keV, requires that the electron temperature is several tens higher, $kT\sim 3$--4 MeV. Then, the origin of the $\sim$100 keV cutoff is not clear; if it is due to the 1st order scattering by those thermal electrons, very strong fine tuning is required to achieve the observed cutoff energy. On the other hand, higher order scatterings will yield another spectral cutoff at $\ga 3kT\sim 10$ MeV, which presence remains to be observationally tested. 

There are two main different models of the hot corona. One relies on electron heating by magnetic reconnection above the standard thin disk (Liang \& Price
1977; Galeev, Rosner, \& Vaiana 1979). Since the process of magnetic reconnection remains poorly understood, this model has not yet been elaborated in detail, although recent numerical simulations have shed some light on this problem (Hirose, Krolik \& Stone 2006). The other model is a hot accretion flow, which is advection dominated (ADAF) in most of its parameter space (Narayan \& Yi 1994, 1995; Abramowicz et al.\ 1995; Rees et al.\ 1982). We need to illustrate two points here when we use the term ADAF. First, it is found that in most of the parameter space of ADAF, outflow is moderately strong,
as emphasized by Blandford \& Begelman (1999, the ADIOS solution) and other authors. Second, as we will show later, another hot accretion solution may be responsible for the X-ray emission of the relatively luminous hard states. 
In comparison with the magnetic reconnection, ADAF has clear dynamics, and its astrophysical applications can be worked out in detail. The specific suggestion of applying the ADAF to the hard state is due to Narayan (1996) and Narayan, McClintock \& Yi (1996) (see also Shapiro, Lightman \& Eardley 1976, who, however, did not recognize the importance of advection). First detailed calculations applied to data were then done by Esin, McClintock \& Narayan (1997, hereafter E97). 

In the present paper we concentrate on the hot accretion flow. We note that the radiative efficiency of an ADAF increases with the accretion rate, $\dot M$. Thus an ADAF can explain not only very dim sources like the Galactic center (e.g., Yuan, Quataert \& Narayan 2003), but also some relatively luminous sources such as hard states of BHXBs. The highest rate, called the critical accretion rate, of an ADAF is determined by the balance between the viscous heating and cooling, $q_{\rm vis}\approx q_{\rm ie}$, which occurs at  $\dot{M}_{\rm ADAF} \sim 10\alpha^2 \dot{M}_{\rm E}$, where $\alpha$ is the viscosity parameter, $\dot{M}_{\rm E} \equiv L_{\rm E}/c^2$, $q$ is the rate of energy change per unit volume, and $L_{\rm E}$ is the Eddington luminosity. The corresponding maximum luminosity is $L_{\rm ADAF} \sim 0.4\alpha^2 L_{\rm E}$ (E97)\footnote{Thus the highest radiative efficiency of an ADAF is $\simeq 0.04$, not much lower than that of a standard thin disk around a Schwarzschild black hole, $\simeq 0.057$.}. The actual value of $\alpha$ remains uncertain, but recent three-dimensional magnetohydrodynamic (MHD) numerical simulations suggest $\alpha\sim 10^{-2}$--$10^{-1}$ in most of the disk midplane (Hawley \& Krolik 2001). Even for $\alpha$ as high as 0.3 (a common value used in ADAF studies), the maximum ADAF luminosity is still only $\simeq (3$--$4)\%L_{\rm E}$. It has been thought that above the $L_{\rm ADAF}$ there is no hot accretion solution, with the thin disk remaining the only viable solution. 

However, observations of the hard state in BHXBs often show higher luminosities during the rising part of an outburst. For example, the hard state of XTE J1550--564 reached $20\%L_{\rm E}$ (Sobczak et al.\ 2000) and $\sim 10\%L_{\rm E}$ (Rodriguez, Corbel \& Tomsick 2003; this work) during its outbursts in 1998 and 2000, respectively. The highest hard-state luminosity of GX 339--4 was $\simeq  25$--$30\%L_{\rm E}$ (Zdziarski et al. 2004; Done \& Gierli\'nski 2003). These observational results have long presented a puzzle for the hot accretion flow models. 

Of importance in this context was the finding of a new hot accretion flow solution, the so-called luminous hot accretion flow (LHAF; Yuan 2001, hereafter Y01). Different from an ADAF, this solution corresponds to accretion rates above
the critical rate of ADAF, $\dot{M}_{\rm ADAF}$, up to another critical 
accretion rate which is about 3-5 times higher than $\dot{M}_{\rm ADAF}$ for
parameters adopted in Y01. In addition to accretion rates, the radiative 
efficiency of an LHAF is also higher than that of a typical ADAF. 
Therefore, an LHAF can produce much higher luminosity than an ADAF. 
Yuan \& Zdziarski (2004) studied some luminous hard states of BHXBs and 
AGNs and found that their relatively low electron temperatures and high 
luminosities do require an LHAF to be present. In the first part of this work, 
we apply this approach to hard-state spectra of XTE J1550--564 during
its 2000 outburst.

In addition to studying X-ray spectra alone, more and more efforts are devoted to multi-wavelength simultaneous observations and modeling of BHXBs, as illustrated, e.g., by studies of XTE J1118+480. Its nearly simultaneous observations were conducted from radio to X-ray wavelengths (e.g., Hynes et al.\ 2000; McClintock et al.\ 2001; Frontera et al.\ 2001, 2003; Chaty et al.\ 2003). These have allowed important theoretical conclusions to be drawn. Esin et al.\ (2001) showed that the EUV data rule out models with a thin disk extending to the last stable orbit. Using a coupled jet-ADAF model, Yuan, Cui \& Narayan (2005) and Malzac, Merloni \& Fabian (2004) successfully explained the radio to X-ray spectrum and most of the timing properties of the source.

Here, we apply the same approach to XTE J1550--564. We use simultaneous multi-wavelength observations at radio, infrared, optical and X-rays during the 2000 outburst (Corbel et al.\ 2001; Jain et al.\ 2001, hereafter J01; Tomsick, Corbel \& Kaaret 2001; Rodriguez et al.\ 2003). Compared to XTE J1118+480, XTE J1550--564 is much more luminous. During the outburst, transitions between the hard and very high states occurred, unlike the case of XTE J1118+480, which was always in the hard state. Therefore some new interesting phenomena were observed. One issue we discuss below is that of secondary maxima in the infrared/optical light curves. J01 reported infrared ($H$), optical ($I$, $V$) and X-ray light curves for the 2000 outburst. After reaching the peak of the outburst, the IR, optical, and X-ray fluxes declined, and a transition from the very high state to the hard state occurred around MJD 51,680. Afterwards, the X-ray flux continued to decline whereas the IR/optical light curves began to rise and reached secondary maxima at MJD 51,699--51,705, after that they declined again. A contemporaneous spectrum in the radio band was obtained by Corbel et al.\ (2001) at MJD 51,697. Here, we model the simultaneous radio, IR, optical, and X-ray data and explain the secondary IR/optical maxima.

In \S \ref{hot_flow}, we model the X-ray emission of the hard states, paying special attention to the dynamics of the hot accretion flow. In \S \ref{jet}, we study the simultaneous radio, IR/optical and X-ray data during the decline phase and explain the secondary IR/optical maxima. We conclude in \S \ref{sum} with a summary and discussion.  

\section{Modeling the X-ray spectra of the hard states of XTE J1550--564}
\label{hot_flow}

\subsection{Observational data}
\label{data}

During its 2000 outburst, XTE J1550--564 experienced hard and very high states. We have selected four occurences of the hard states, at MJD 51,646, 51,658, 51,687, and 51,696, based on the criteria that they span a wide range of X-ray luminosity and that each includes simultaneous optical data. Their spectra are shown in Figure \ref{spectra} by different colors. For comparison, we also show an example of the very high state spectrum, of MJD 51,662. Their X-ray \xte\/ spectra were earlier studied by Rodriguez et al.\ (2003) and Tomsick et al.\ (2001); here we have reextracted them ourselves. They can be well fitted by a power-law form with an exponential cut-off. The X-ray spectra in blue, red, green, and magenta in Figure  \ref{spectra} (from top to bottom) are fitted with the photon index and the e-folding energy, $(\Gamma, E_{\rm f})$, of $(1.70\pm 0.01, 115\pm 5.6), (1.46\pm 0.01, 137^{+8}_{-7}), (1.53\pm 0.01, 200^{+80}_{-50})$ and $(1.53\pm 0.01, 460^{+340}_{-240}$), respectively. We see a strong anticorrelation between the X-ray flux and $E_{\rm f}$, with its value monotonously increasing with the decreasing flux. 

The IR/optical data in Figure \ref{spectra} are from J01. Since we do not know the exact value of the extinction, we show two sets of fluxes for each spectrum, corresponding to the two limits of the extinction of $A_V=4.75$ (Orosz et al.\ 2001) and $A_V=2.2$ (S{\'a}nchez-Fern{\'a}ndez et al.\ 1999). The details of the data reduction can be found in Xue, Wu, \& Cui (2007). 

\subsection{Accretion flow models}
\label{models}

We now fit the spectra using our accretion flow model. The components of it are nearly the same as in E97, i.e., the accretion flow consists of an inner hot accretion flow (ADAF or LHAF) within a transition radius, $r_{\rm tr}$, and an outer thin disk. We assume that only a fraction of the accretion rate at $r_{\rm tr}$ actually accretes onto the black hole and the rest is ejected. This assumption is required by both numerical simulations (Stone, Pringle, \& Begelman 1999; Hawley \& Balbus 2002; Igumenshchev et al.\ 2003) and analytical work (Narayan \& Yi 1994; Blandford \& Begelman 1999; Narayan et al.\ 2000; Quataert \& Gruzinov 2000). Details of this process depend on the accretion rate. Here, in order to formally take into account the role of outflows as well as convection in modifying the density profile of the accretion flow, we assume following Yuan et al.\ (2005), that $\dot M(r)=\dot M_0$ for $r\geq r_{\rm tr}$, and 
\be 
\frac{{\rm d} \ln \dot{M}(r)}{{\rm d} \ln r} = s(r), \qquad r< r_{\rm tr},
\label{mdot}
\ee
where
\be s(r) = s_0 \max[f(r),0], 
\label{s_r}
\ee
and $s_0$ is independent of $r$ but it can be different for different values of $\dot M_0$. The above formula is based on the consideration that the outflow (and convection) is ultimately due to the accreting gas acquiring a positive Bernoulli parameter, as emphasized by Narayan \& Yi (1994), and which value is smaller for higher accretion rates due to the stronger radiative energy loss. Then, $f(r)$ is the advection factor of the accretion flow, defined as, 
\be f(r) \equiv \frac{q_{\rm adv}}{q_{\rm vis}}= \frac{q_{\rm vis}-q_{\rm ie}}
{q_{\rm vis}}. 
\label{f_r}
\ee 
A negative value of $f$ in eq.\ (\ref{f_r}) implies that advection plays a heating rather than a cooling role. In this case, the hot accretion flow is in the LHAF regime (Y01), as 
we will point out in \S 2.3. When $\dot M$ is very low, $f(r)=1$ and $s(r)=s_0$. In this case, eq.\ (\ref{mdot}) gives us the usual form, $\dot{M} = \dot{M}_0 (r/r_{\rm tr})^{s_0}$ (e.g., Blandford \& Begelman 1999). 

We calculate global solutions of the hot accretion flow using the method of Y01.  A major difference of that method with respect to that of E97 is that we solve the radiation hydrodynamics equations self-consistently, and thus obtain $f(r)$ at each radius.  In contrast, E97 used the approximation of $f(r)$ having a constant average value at all radii. The radiation processes we consider include bremsstrahlung, synchrotron emission, and Comptonization of both synchrotron photons from the hot accretion flow and soft photons from the thin disk. We use formulae in Narayan \& Yi (1995) and Coppi \& Blandford (1990) to calculate the synchrotron and Comptonization spectra, respectively, as a function of the radius. We treat self-absorption as in Manmoto, Mineshige \& Kusunose (1997). The emission from the outer cool disk is modeled as a multicolor blackbody spectrum. The temperature as a function of $r$ is determined by the viscous dissipation and the irradiation of the disk by the inner hot flow. E97 show that when $r_{\rm tr}\ga 30 r_{\rm S}$, the outer thin disk has little effect to the X-ray emission from the inner hot accretion flow because the seed photons for the Comptonization mainly come from the synchrotron and bremsstrahlung emissions in the hot accretion flow. Here $r_{\rm S}\equiv 2GM/c^2$ is the Schwarzschild radius. 

Now we discuss the values of the model parameters. As we stated in \S \ref{intro}, the exact value of $\alpha$ is uncertain, and we adopt $\alpha= 0.3$, which appears the highest possible value. This choice allows us to show the highest values of $L$ possible with the ADAF model, and it shows that even then that model cannot explain the luminous hard states of BHXBs. We set the magnetic parameter, $\beta$ defined as the ratio of the gas pressure to the sum of the gas and magnetic pressures, at $\beta=0.9$ (from MHD numerical simulations; e.g., Hawley \& Krolik 2001), and the fraction of the viscous dissipation directly heating electrons is set at $\delta=0.5$ (from the modeling to the supermassive black holein our Galactic center: Yuan et al.\ 2003). The parameters $\alpha$, $\beta$ and $\delta$ are not free hereafter, though we realize that their values bear large uncertainties. The outer radius of the cold disk is set equal to the tidal radius, which is $\sim$0.87 of the Roche lobe radius (Papaloizou \& Pringle 1977). The Roche lobe radius is 0.55--0.59 of the separation (Eggleton 1983) for the estimated mass ratio of the black hole to the companion of 6.7--11, while the separation is $\simeq 9\times 10^{11}$ cm (Orosz et al.\ 2002). This implies the outer radius $\simeq 5\times 10^{11}$ cm, equivalent to $\simeq 1.5\times 10^5(10\msun/M) r_{\rm S}$. We set the black-hole mass to $M=10.5 \msun$ and the distance to $D=5.3$ kpc (Orosz et al.\ 2002)\footnote{Jonker \& Nelemans (2004) argued that the distance to XTE J1550--564 might have been underestimated by as much as a factor of three due to an overestimation of the interstellar extinction. However, if it were true, the  highest 1--20 keV luminosity of the 1998 outburst would be $\sim 10L_{\rm E}$, which seems highly unlikely. }. 

Another important parameter is the transition radius, $r_{\rm tr}$. As it has been shown by Yuan et al.\ (2005), its value is best constrained by the thin disk emission close to $r_{\rm tr}$, which is in the EUV/soft X-ray band. 
However, we have no data for that band, and hereafter assume $r_{\rm tr}=100 r_{\rm S}$. As stated above, most of the seed photons for Comptonization at this value of $r_{\rm tr}$ are synchrotron photons from the hot accretion flow. Our satisfactory fits to the data (the magenta, green and red colors in Figure \ref{spectra}) indicate that our solution is viable though it may be not unique. 

On the other hand, it is possible that when the hard state approaches the very high state, the soft photons from the thin disk or clumps in the hot accretion flow become dominant (see also Wardzi\'nski \& Zdziarski 2000). As it has been shown by Yuan et al.\ (2007), the X-ray slope of (at least some) black-hole binaries is correlated with the luminosity, $L$, and the sign of the correlation changes when the luminosity of the hard state increases above a certain value. Below this luminosity, the X-ray spectrum hardens with the increasing $L$, and above it, it softens. The boundary in our case approximately corresponds to the red-color spectrum in Figure \ref{spectra}, though in our case with a limited range of $L$ the X-ray spectra below it keep an approximately constant slope. The change of the sign of the correlation may be due to the change of the dominant source of the seed photons.

\subsection{Modeling results}
\label{results}

Figure \ref{spectra} shows our X-ray spectral fits to the lower three hard states. (We will discuss the brightest hard state separately.) From bottom to top, they are shown by the dot-dashed, dashed and solid curves and have $(\dot M_0/\dot{M}_{\rm E}, s_0)$ equal to $(1.0, 0.55)$, $(1.1, 0.3)$, and $(1.3, 0.3)$, respectively. The advection factor, $f$, and the electron temperature, 
$T_{\rm e}$, of these models are shown as a function of $r$ in Figures 
\ref{advection} and \ref{T_e}, respectively.

We see in Figure \ref{spectra} that the three models fit the X-ray spectra 
relatively well, including both the spectral slope and the cut-off energy. 
E97 did not take into account outflows and predicted that the X-ray spectra 
from hot accretion flows should harden with the increasing accretion rates. 
The X-ray spectrum at MJD 51,646 (in red) is harder than that at MJD 51,687 
(in green), consistent with that prediction. However, the slope of the 
spectrum at MJD 51,687 is identical to that at MJD 51,696 (magenta), 
different from the prediction of E97. Our calculations show that the 
inclusion of the outflow in our model can mostly resolve this discrepancy. From eqs.\ (\ref{mdot}--\ref{s_r}) and Figure \ref{advection}, we see that there is no outflow for the model shown by the (high) solid curve since its advection factor is $f<0$. For the (middle) dashed line model, $f$ is very small: 
$f\approx 0$, so outflow is also weak. Since the accretion rate of the solid curve is larger than that of the dashed curve, the former spectrum is harder. On the other hand, for the (low) dot-dashed curve, $s_0=0.55$ and $f\sim 1$, the outflow is strong. 

Then, the presence of an outflow will make the emitted spectrum harder 
compared to that with no outflow at the same $\dot{M}_0$. For comparison, 
we also show in Figure \ref{spectra} the spectrum produced by 
a model without outflow (at $\dot{M}_0=0.5\dot{M}_{\rm E}$ to reproduce the X-ray flux level), see the double-dot-dashed curve. We clearly see that its spectrum is softer than that of the dot-dashed curve (including the outflow). The physical reason for this is as follows. The emitted X-ray spectrum is the sum of the local Comptonization spectra at each radius, and the slope of that spectrum is determined by the Compton $y$ parameter, with a larger $y$ corresponding to a harder spectrum (e.g., Beloborodov 1999). At small radii, $y$ is relatively small so the corresponding spectrum is soft. When a strong 
outflow exists, the accretion rate decreases with decreasing radii, 
thus the contribution of the innermost region (where the dissipation 
peaks) becomes smaller. Therefore, the total spectrum will become 
harder. We note that this effect is not universal; it holds only 
for a certain range of the accretion rate.

We note that the spectra produced by all the three models are somewhat too soft in comparison with the X-ray data. Increasing slightly $\dot{M}_0$ can make the spectrum harder, thus improving the fits. However, the corresponding 
luminosities will be higher than observed values. Given the uncertainty in the distance of XTE J1550--564, it is possible that the distance is slightly larger than $D=5.3$ kpc. Another solution would be to use a larger $\beta$, which will reduce the magnetic field and thus the synchrotron emission. As the latter provides seed photons for Comptonization, this will lead to hardening of the spectra. 
 
Our models also fit satisfactorily the e-folding energies, $E_{\rm f}$. As noted in \S\ref{data}, , $E_{\rm f}$ is anticorrelated with the X-ray luminosity. Such an anticorrelation appears common in the hard states. It takes place, e.g., 
in GRO J0422+32 (Grove et al.\ 1998) and GX 339--4 (Wardzi{\'n}ski 
et al.\ 2002). In thermal Comptonization models, $E_{\rm f}$ is close to 
the electron temperature $T_{\rm e}$. Such an anticorrelation between 
$T_{\rm e}$ and $L$ (or $\dot M$) is expected in the hot accretion 
flow models, both ADAF and LHAF, as shown here by Figure \ref{T_e} (see also Fig.\ 3b of E97). It shows the $T_{\rm e}$ profiles of the three models shown in 
Figure \ref{spectra}, together with the values of $E_{\rm f}$ of the observed 
X-ray spectra. Note the value of $E_{\rm f}$ for the lowest considered state 
is not precisely determined. Since the emitted X-ray spectra are integrated 
over the radii, the $E_{\rm f}$ here only measures an average $T_{\rm e}$. 
Still, the shown agreement is excellent, indicating that the hot accretion flow model is able to quantitatively predict the correct $E_{\rm f}$. We emphasize that this result is quite robust, independent of the model parameters. It provides strong support for this model of the hard state.

Let us now discuss the dynamics of the accretion flow. We first need to 
illustrate the physics of the LHAF. The energy equation of ions in hot 
accretion flows reads: $q_{\rm adv}\equiv q_{\rm int}- q_{
\rm comp}=q_{\rm vis}-q_{\rm ie}$. The subscripts denote the energy advection, 
the gradient of the internal energy of ions, compression work, viscous heating, 
and Coulomb collision cooling, respectively. The critical accretion rate of an ADAF $\dot{M}_{\rm ADAF}$ is determined by $q_{\rm vis} = q_{\rm ie}$ 
(Narayan, Mahadevan \& Quataert 1998). Below $\dot{M}_{\rm ADAF}$, 
$q_{\rm vis} > q_{\rm ie}$, therefore we have 
$q_{\rm int}= q_{\rm comp} +q_{\rm vis} -q_{\rm ie} > 0$. {\em The positive sign of 
$q_{\rm int}$ means that the flow can remain hot if it starts out hot 
(because $q_{\rm int}\propto v dT/dr$, and $v<0$). This is why a hot 
ADAF solution can exist}. Above $\dot{M}_{\rm ADAF}$ 
up to another (higher) critical accretion rate determined 
by $q_{\rm vis} + q_{\rm comp} = q_{\rm ie}$,
the value of $q_{\rm int} (= q_{\rm comp} +q_{\rm vis} -q_{\rm ie})$ is still
positive. Thus a new hot accretion solution, i.e., LHAF,
exists between $\dot{M}_{\rm ADAF}$ and this new rate (Y01). For an
ADAF, $q_{\rm adv}>0$, so advection plays a cooling role. While in an LHAF,
$q_{\rm adv} <0$, so advection plays a heating role. In this case, 
the radiative cooling is balanced by the sum of advective and viscous
heating.

In Figure \ref{advection}, we see that the advection factor for the (low) dot-dashed 
curve is $>0$, thus it represents an ADAF. The bolometric luminosity of this model is 
$L\sim 0.8\% L_{\rm E}$. The advection factor for the (middle) dashed curve is negative in some region, and thus it is marginally an LHAF, 
with $L\sim 3\%L_{\rm E}$. This result is roughly consistent with those of E97, who found the highest accretion rate of ADAFs as $\dot{M}_{\rm ADAF} \approx 13 \alpha^2 \dot{M}_{\rm E} \approx \dot{M}_{\rm E}$ and the corresponding 
highest $L$ is $\simeq 0.4\alpha^2 L_{\rm E} \approx (3$--$4)\%L_{\rm E}$ 
for $\alpha=0.3$. Thus, for spectra such as that of MJD 51,687, 
whose $L$ is $\sim 3\% L_{\rm E}$, an ADAF marginally does not work. 
For the spectrum at MJD 51,646, the luminosity is still higher, 
$\sim 6\%L_{\rm E}$. In this case, the flow is an LHAF in the entire range, see the solid curve in Figure \ref{advection}. 

However, even an LHAF cannot explain the spectrum at MJD 51,658 (the blue 
symbols).  This is because the required $L$ is above the maximum 
accretion rate of LHAF, $\dot{M}_{\rm LHAF}$, roughly 
determined by $q_{\rm vis}+q_{\rm comp}=q_{\rm ie}$. Because $q_{\rm comp}$ 
does not depend on $\alpha$, the value of $\dot{M}_{\rm LHAF}$ depends only 
weakly on $\alpha$, unlike $\dot{M}_{\rm ADAF}$. In addition to $\alpha$, 
other model parameters such as $\beta$, $\delta$ and the outer boundary 
conditions also affect its value. Y01 found that 
$\dot{M}_{\rm LHAF}/\dot{M}_{\rm E}\simeq 5$ and $3$ for $\alpha=0.3$ and $0.1$, 
respectively. For the parameters adopted here, $\dot{M}_{\rm LHAF}$ is slightly 
larger than $1.3\dot{M}_{\rm E}$\footnote{As a comparison, the critical rate 
of the ADAF for the parameters adopted in the present paper is 
$\dot{M}_{\rm ADAF} \simeq 1.1\dot{M}_{\rm E}$. The 
$\dot{M}_{\rm LHAF}/\dot{M}_{\rm ADAF}$ ratio is smaller than that 
obtained in Y01. This is due to the different parameters assumed, 
$\alpha, \beta$ and $\delta$ as well due to the inclusion of outflows 
in the ADAF but not in the LHAF (where they are suppressed).}, 
which corresponds to $L\la 8\% L_{\rm E}$. This is not high enough to 
explain the X-ray spectrum at MJD 51,658. When $\dot{M}_0\ga \dot{M}_{\rm LHAF}$, the innermost region of the LHAF will collapse and form an cold annulus (Y01). Calculations of the emitted spectrum in this case are complicated; here, we hypothesize that this effect may be responsible for the very high state.

Phenomenologically, the X-ray spectrum at MJD 51,658 appears to be intermediate between the hard state spectra, and the very high state spectrum, shown by the black points. The approximate agreement between the observed highest luminosity of the proper hard state and the highest LHAF luminosity strongly supports the LHAF model for the hard state of BHXBs. 

Figure \ref{spectra} shows that the IR/optical fluxes are under-predicted 
by the hot flow models. This implies that some other components such as a 
jet and the thin disk also contribute. We consider the role of the jet 
in \S \ref{jet}. Here, we show the emission from an irradiated truncated 
thin disk with $\dot{M}= 1.3\dot{M}_{\rm E}$ and $r_{\rm tr}=100 r_{\rm S}$. 
We see that this improves the agreement of the model with the data in the optical band, but most of the IR data are still not explained. 

\section{The role of jet}
\label{jet}

We note here two interesting results in the IR/optical and X-ray light 
curves for the 2000 outburst reported by J01. The first one is that the 
outburst in the IR/optical started some days before that in X-rays 
(seen by the \xte/ASM). Here, we focus on another interesting result of J01. 
After the primary peak of the outburst, the $HIV$ and X-ray fluxed declined, and the transition from the very high state back to the hard state occurred around MJD 51,680. After the transition, the X-ray flux continued to decline while the $HIV$ fluxes began to rise and reached secondary maxima at MJD 51,699--51,705. J01 also found that the secondary maximum is most prominent in the $H$ band. 

\begin{figure} \epsscale{0.7} \plotone{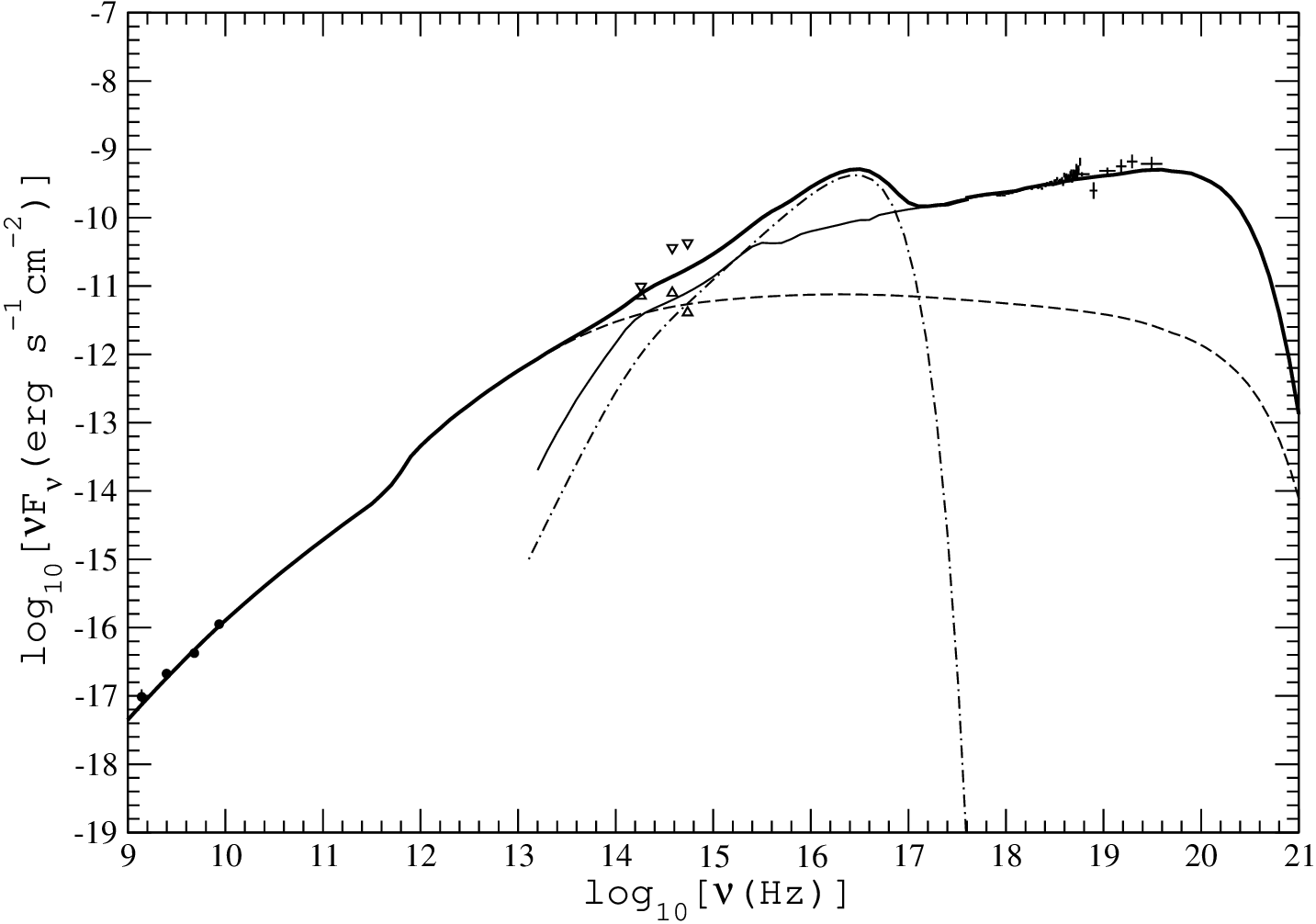} \vspace{.4in}
\caption{An accretion-jet model for XTE J1550--564 at MJD 51,696. 
The dashed, thin solid and dot-dashed curves show the emission from the jet, the LHAF, and the truncated thin disk, respectively. The jet emission contributes significantly at the $H$, $I$, and $V$ bands. On the other hand, the jet contribution is negligible in the X-ray band. This explains why we see secondary peaks at the $H$, $I$, and $V$ bands with no corresponding peak at the X-ray band. See \S \ref{jet} for details.
\vspace{.0in}
\label{jet_spectrum}}
\end{figure}

The IR/optical and X-ray spectra at MJD 51,696 are shown in Figure \ref{jet_spectrum}. It also shows the contemporaneous radio measurements at MJD 51,697 (Corbel et al.\ 2001) with the Australia Telescope Compact Array (ATCA) at four frequencies from 1384 to 8640 MHz. The detected spectrum is slightly inverted with the spectral index of $\alpha_{\rm R} =0.37\pm 0.10$ (where the energy flux is $F_\nu\propto \nu^{\alpha_{\rm R}}$). Since the radio regime is much less variable than the X-rays, we assume that the radio spectra at 
MJD 51,696 and 51,697 were the same.

Based on Figure \ref{jet_spectrum}, we can rule out the standard thin disk as the origin of the secondary $HIV$ peaks. The flux produced by the thin disk in the $H$ band depends mainly on the values of the outer radius of the thin disk, $r_{\rm out}$, and $\dot{M}$. The dot-dashed curve in Figure \ref{jet_spectrum}
shows the emitted spectrum by an irradiated thin disk with $\dot{M}_0= 1.0 \dot{M}_{\rm E}$ and $r_{\rm out}=1.5\times 10^5 r_{\rm S}$. We can see that the emission from the disk (in the Rayleigh-Jeans regime) is substantially below the observed $H$ flux. We note that only for $\dot M \sim 40 \dot{M}_{\rm E}$, 
which appears highly unlikely, the observed and predicted $H$ fluxes would become equal. 

Jet formation is usually observed when the source enters the hard state (Fender 2001, 2006). Prompted by this, J01 proposed that the secondary peaks are due to the emission from the jet formed at the time of the very high-hard state transition around MJD 51,680. Our detailed calculations support this conjecture.
We model the MJD 51,696 spectrum (Figure \ref{jet_spectrum}) taking into account
the contributions from the hot flow, the truncated disk, and the jet. We use the internal shock scenario (widely adopted in the study of GRB afterglows) to calculate the jet emission, see Yuan et al.\ (2005) for details. Briefly, internal shocks within the jet occur due to collisions of shells with different velocities. The shocks accelerate a fraction of the electrons into a nonthermal, power-law, energy distribution. The steady-state electron energy distribution is then self-consistently determined taking into account the radiative cooling. We parametrize the model by two parameters, $\epsilon_{\rm e}$ and $\epsilon_B$, giving the fraction of the shock energy going into the accelerated electrons and the (amplified) magnetic field, respectively. We then calculate the nonthermal synchrotron emission. 

The modeled jet and total spectra are shown by the dashed and solid curves 
respectively in Figure \ref{jet_spectrum}. We can see that this model fits the data from the radio to the X-rays very well. The jet parameters are: $\epsilon_{\rm e} =0.06$, $\epsilon_B =0.02$, its opening angle of $\theta=0.1$, the bulk Lorentz factor, $\Gamma_{\rm jet}= 1.2$, and the spectral index at which the electrons are accelerated of $p=2.23$. The values of the parameters are the same as those used by Yuan et al.\ (2005) to model the outburst hard state of XTE J1118+480. The mass flow through the jet is $\dot{M}_{\rm jet}=3.3\times 10^{-3}\dot{M}_{\rm E}$. This is $\sim 0.6\%$ of the accretion rate at in the inner flow, equal to $0.54\dot M_{\rm E}$ at $5 r_{\rm S}$.
This is very similar to the corresponding value of $0.5\%$ in XTE J1118+480. 

Based on the above fitting, we now explain the secondary $HIV$ peaks as follows. The jet began to form at the very-high to hard state transition, at MJD 51,680. Initially, $\dot{M}_{\rm jet}$ was very small, therefore the contribution of the jet emission to the $H$ flux was negligible. But then the accretion $\dot{M}$ continued to decrease and the accretion luminosity declined. On the other hand, the jet grew, with $\dot{M}_{\rm jet}$ increasing, so the relative jet contribution to the $HIV$ bands continued to increase. The increase of $\dot{M}_{\rm jet}$ lasted for $\sim$20 d and reached its maximum at MJD 51,699--51,705, which corresponds to the secondary peaks in the $HIV$ light curves. After that time, $\dot{M}_{\rm jet}$ decreased together with the $\dot{M}$. Our fitting results at MJD 51,696, shown in Figure \ref{jet_spectrum}, are for the time very close to that of the secondary $HIV$ peaks.

Fig.\ 1 in J01 shows that the $H$ band fluxes at MJD 51,680 and 51,696 differ by one magnitude. We note that the modeled ADAF and the thin disk fluxes could have changed only very little between those two days (see Figure \ref{spectra}). On the other hand, from Figure \ref{jet_spectrum}, we see that the $H$ band flux increases due to the appearance of the jet emission at MJD 51,696 by about one magnitude, completely consistent with the observational results in J01. 
We also see in Figure \ref{jet_spectrum} that while the $\nu F_\nu$ spectrum 
from the jet is almost flat in the $HIV$ band, the spectrum of the accretion flow (the thin disk and ADAF) rises with the increasing frequency. This explains why the secondary peak is most prominent at the $H$ band (J01), i.e., at the lowest energy.

Another important observational result is the absence of the corresponding secondary peak in the X-ray light curve. This requires that whatever emission the jet produces in X-rays, it has to be negligible compared to that
from the hot accretion flow. Our theoretical modeling is consistent with that
result. As shown in Figure \ref{jet_spectrum}, the contribution of the jet emission in the X-ray band is much weaker than that of the accretion flow. We thus conclude that the absence of the secondary maximum in the X-ray light curve provides a strong evidence that the jet emission (synchrotron plus Compton) is not important in the X-ray band in the hard state. 

\section{Summary and Discussion}
\label{sum}

\begin{figure} \epsscale{0.7} \plotone{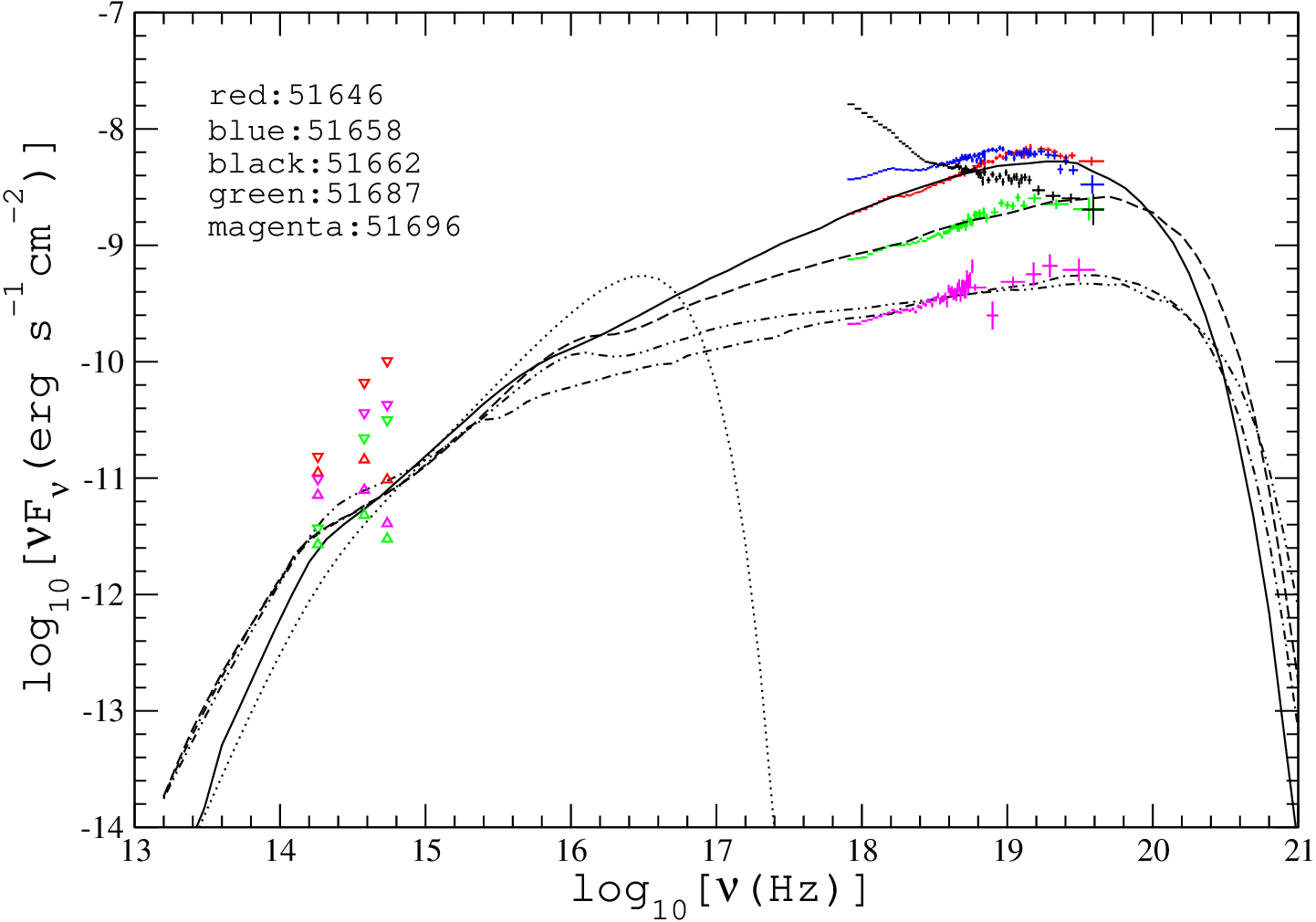} \vspace{.2in}
\caption{Spectral fitting results for the hard states of XTE J1550--564 during its 2000 outburst at different dates. The two sets of optical data for each date correspond to the two limiting values of the extinction, $A_V=4.75$, 2.2. The 
parameters of the hot accretion flow models are: $\dot{M}_0= 1.0\dot{M}_{\rm E}$ and $s_0=0.55$ (dot-dashed curve); $\dot{M}_0=1.1\dot{M}_{\rm E}$ and $s_0=0.3$ (dashed curve); and $\dot{M}_0=1.3\dot{M}_{\rm E}$ and $s_0=0.3$ (solid curve). The dot-dashed curve shows an ADAF model while the other two models are LHAFs (see Figure \ref{advection}). The double-dot-dashed curve shows a model without an outflow to illustrate its hardening effect on the X-ray slope (see \S \ref{results}). The dotted curve shows the emission from a truncated thin disk at $\dot{M}= 1.3\dot{M}_{\rm E}$ and $r_{\rm tr}=100 r_{\rm S}$. We see that the inclusion of this component imroves the fit in the optical range, but it is not sufficient to explain the IR data. 
\vspace{.4in}
\label{spectra}} 
\end{figure}

\begin{figure} \epsscale{0.7} \plotone{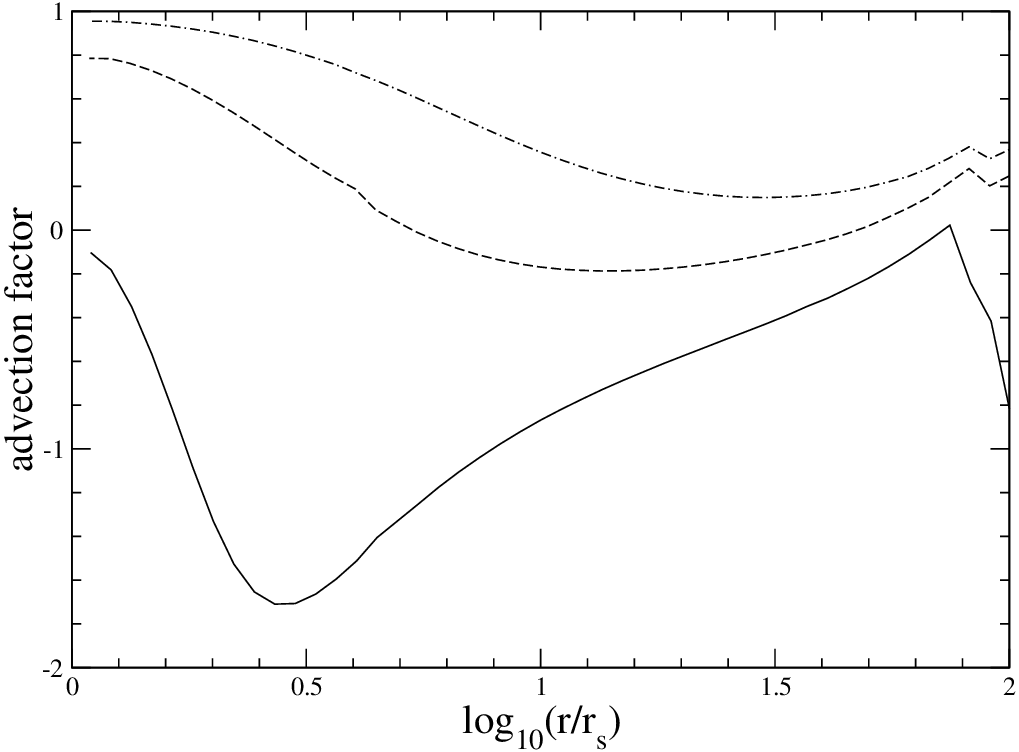} \vspace{.40in}
\caption{The profile of the advection factor (defined in eq.\ \ref{advection})
for the three models shown in Figure \ref{spectra}. The model denoted by the dot-dashed curve is an ADAF since its advection factor is always positive, while
the other two models are LHAFs since their advection factor are negative in some regions. The irregular behavior near 100$r_{\rm S}$ is due to an effect of the boundary condition.
\label{advection}} 
\end{figure}

\begin{figure} \epsscale{0.7} \plotone{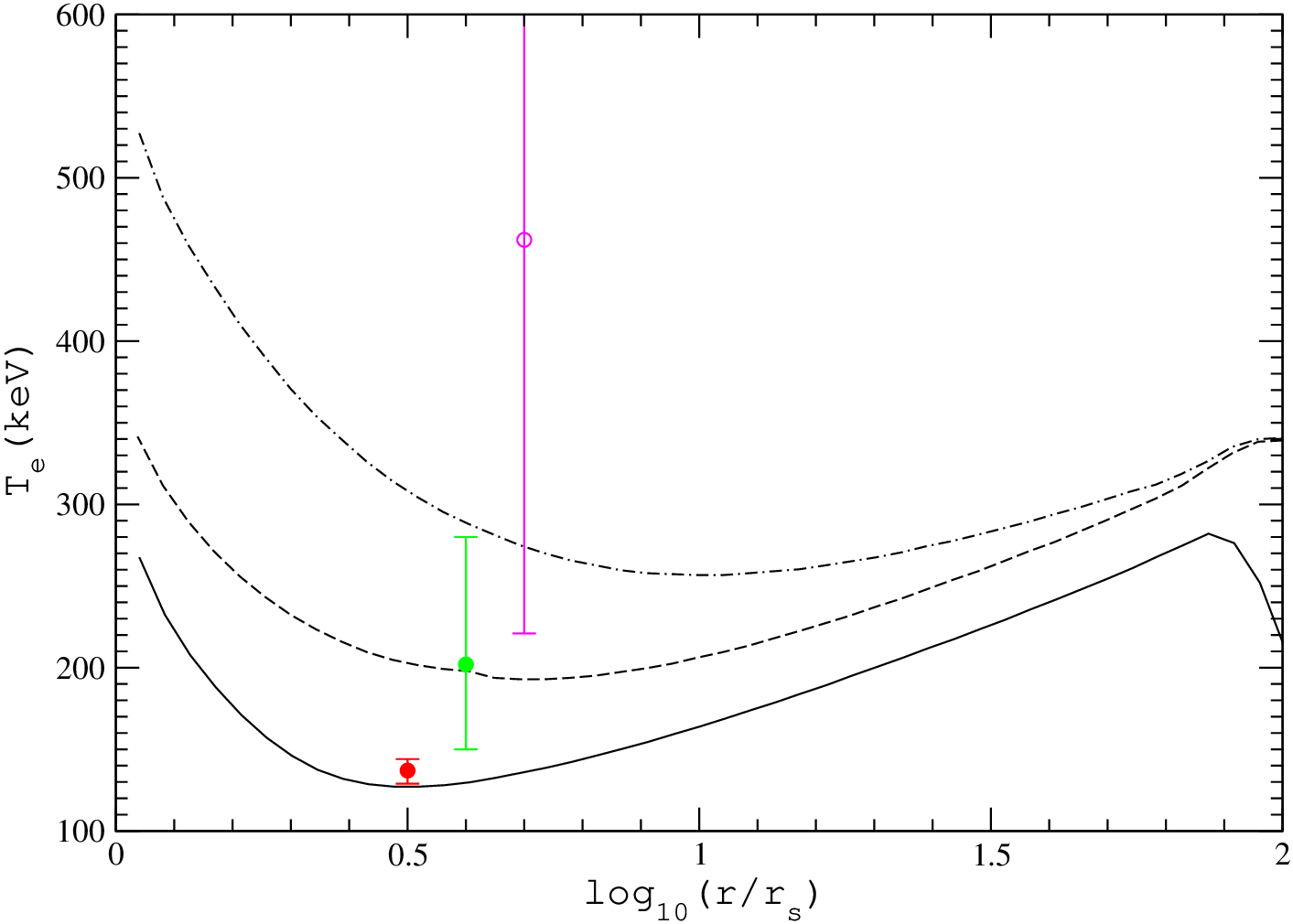} \vspace{.4in}
\caption{The profile of the electron temperature for the three models
shown in Figure \ref{spectra}. With the increase of the accretion rates, the
electron temperature decreases, consistent with the observed
anticorrelation between the e-folding energy, $E_{\rm f}$, and the luminosity seen in Figure \ref{spectra}. The three circles with different colors
show the $E_{\rm f}$ for the corresponding three X-ray spectra shown in Figure \ref{spectra}. Their abscissa is arbitrary. The magenta circle (open) denotes the e-folding energy poorly constrained by the data. Since $E_{\rm f}$
is proportional to the electron temperature, these results show
that the hot accretion flow models (ADAF and LHAF) can quantitatively
predict the correct electron temperature.
\label{T_e}} 
\end{figure}

We have investigated the hard states of XTE J1550--564 during its 2000 outburst. We have modeled the X-ray spectra with accretion flow models (\S \ref{hot_flow}). Our model consists of an inner hot flow within a transition radius and an outer thin disk, as in E97. Different from E97, we have taken into account two new developments of the hot accretion flow models. One is the inclusion of outflow, the other is the extension of the ADAF model to higher accretion rates using the LHAF model (Y01). Our model fits well the lower three of the four characteristic hard-state spectra, see Figure \ref{spectra}. The two extensions of the model are necessary for the obtained good fits. First, the inclusion of the outflow allows us to fit the X-ray spectral slopes. Without outflow, the emitted spectrum hardens with the increasing accretion rate (E97), which disagrees with the observations, where the slopes of the two lowest data sets are  almost identical in spite of their different luminosities. To resolve this discrepancy, the outflow becomes weaker with the increasing accretion rate. This is physically motivated as the radiative energy loss is smaller at lower accretion rates and thus the Bernoulli parameter is larger.

The extension of hot accretion flow models from ADAFs to LHAFs enables us to model the 
luminous hard states with luminosities up to $\sim 8\%L_{\rm E}$ (while the ADAF is 
limited to $\la 3\%L_{\rm E}$). The LHAF fits the spectral slope and 
the cut-off energy of those states, as shown in Figure \ref{spectra}, confirming the prediction of Yuan \& Zdziarski (2004).

The value of the highest possible luminosity of the LHAF model, $L_{\rm LHAF}$,  
depends on $\alpha$ and other model parameters, and the outer boundary condition 
(see Y01). For our assumed parameters, we find $L_{\rm LHAF}\simeq 8\%L_{\rm E}$, somewhat lower than $L$ of the most luminous hard state considered here 
shown by the blue symbols in Figure \ref{spectra}. However, we can see that this state appears intermediate between the hard and the very high states, and perhaps some prcesses responsible for the very high state are already for importance here. 
{\em Thus, the overall agreement between the maximum observed $L$ and the 
prediction of the LHAF model provides strong support for it as a model of the 
X-ray emission of the hard states of BHXBs}. 

The highest $L$ of the hard state varies among various sources as well as different outbursts of a given source. As discussed in \S \ref{intro}, the maximum hard-state $L$ of XTE J1550--564 and GX 339--4 varies in the $\sim$10--30\% range. On the other hand, the corresponding value in Cyg X-1 is only $\sim 0.01$--$0.02 L_{\rm E}(d/2\,{\rm kpc})^2(10\msun/M)$ (e.g., Zdziarski et al.\ 2002). We discuss now possible reasons for this variability. 
 
First, $L_{\rm LHAF}$ is a function of the model parameters, e.g., $\alpha$, and parameter variations among different sources will contribute to its dispersion. However, numerical simulations of accretion flow indicate that $\alpha<0.3$ (e.g., Hawley \& Krolik 2001). Thus, this effect can explain a low value of the maximum $L$, but not values $\ga 10\%L_{\rm E}$. Still, we can obtain $\sim 15\%L_{\rm E}$ by adjusting the outer boundary condition (determined by the physics of the transition region between the hot and cool accretion flows, which is complex and currently unclear). 

An efficient way of increasing the maximum hard-state $L$ can be due to the angular momentum of the black hole being $>0$. In the present calculations, we have assumed a Schwarzschild black hole. If the black hole is rapidly rotating, substantially more gravitational energy will be released because of the deeper potential well (e.g., Gammie \& Popham 1998; Popham \& Gammie 1998). Shafee et al.\ (2006) indeed have found that a number of BHXBs appear to have the spin parameter of $\sim$0.6--0.8.

Yet another effect that could increase $L_{\rm LHAF}$ is a magnetic coupling
between the rotating black hole and the accretion flow. If the black hole rotates faster than the disk, energy and angular momentum will be extracted from the black hole and transferred to the disk by the magnetic field (Li 2000, 2002; Wang et al.\ 2003). However, we are not able to estimate quantitatively the importance of this effect here because the existing studies have only considered the standard thin disk rather than a hot flow\footnote{Recently attention has been paid to the effect of the presence of magnetic stresses near and inside the
last stable circular orbit on the radiative efficiency of the standard thin disk (e.g., Agol \& Krolik 2000; Beckwith, Hawley \& Krolik 2006). This effect does not apply to the hot accretion flow since the no-torque boundary condition is not adopted in the global solution of hot accretion flow.}. 

As we point out above, the most luminous hard states border the very high state. The nature of the latter remains an open question. Numerical calculations by Y01 (see also Begelman, Sikora \& Rees 1987) indicate that above $\dot{M}_{\rm LHAF}$ the hot accretion flow will collapse within a certain radius and form a cold annulus (see Liu, Meyer \& Meyer-Hofmeister 2006 for a similar result in the context of their evaporation model). The inner cold annulus will emit black body radiation which may explain the thermal component of the very high state. At the collapse region, strong magnetic reconnection events should occur due to the overflow of the magnetic field (Begelman et al.\ 1987; Y01). This is likely to accelerate some electrons into a power-law distribution. Compton scattering of the soft photons from the inner annulus by both the thermal and power-law electrons may then be responsible for the X-ray emission of the very high state. We note that Gierli{\'n}ski \& Done (2003) have found that the shape of the soft $\gamma$-ray spectrum of XTE J1550--564 in the very high state indeed requires the presence of both thermal and nonthermal (power-law like) electrons.

A possible problem with the above scenario is that the hot accretion flow 
is strongly thermally unstable at the collapse region (Yuan 2003). As a result of the instability, the accretion flow may form a two-phase flow structure, with cold clumps surrounded by the hot phase (Yuan 2003; Guilbert \& Rees 1988; Celotti, Fabian \& Rees 1992; Kuncic, Celloti \& Rees 1997; Krolik 1998; Beloborodov 1999; Merloni et al.\ 2006). The radiation from the cold clumps can then explain the thermal component of the very high state, whereas Comptonization of the soft photons by the high-energy electrons in the hot phase may be responsible for the power law tail. Thus, this represents an alternative scenario for the very high state. (However, a hybrid electron distribution in the hot phase is required as stated above, see Gierli{\'n}ski \& Done 2003.) We speculate that even some luminous hard-state flows with $\dot{M}\la \dot{M}_{\rm LHAF}$ may also contain such a two-phase configuration. The formation of the cold clumps in this case could be due to the strong instability in the transition region between the hot accretion flow and the standard thin disk (Kato 1999, 2000). This scenario is promising in explaining the observed power spectra of the rapid aperiodic variability of X-ray emission of BHXBs (B\"ottcher \& Liang 1999). It also may explain the claimed relativistic iron K emission line profiles in BHXBs (e.g., Miller et al.\ 2006). We plan to investigate these issues in future work.

We have also investigated the nature of the secondary peaks in the $HIV$ light curves of XTE J1550--564 (J01). They began to emerge after the transition from the very high state back to the hard state, and reached the maximum $\sim$20 d later. However, there have been no corresponding peak in the X-ray light curve. Based on a detailed modeling to the multi-waveband spectrum from the radio to X-rays, we conclude that the secondary peaks are due to the emission from the jet formed during the state transition. The mass loss rate in the jet kept increasing for $\sim$20 d after its birth, and then decreased. The secondary peaks were then due to the corresponding maximum of the relative contribution of the jet ($HIV$) emission with respect to the accretion emission. The absence of the corresponding peak in the X-rays strongly indicates that the jet emission (either synchrotron or Compton) is negligible in the X-ray band in the hard state. This differs from the scenarios with the dominance of the jet in X-rays 
(e.g., Markoff et al.\ 2005). In principle, it is possible that a drop in the X-ray flux from the disk is exactly compensated by a rise in the X-ray flux from the jet, with the total flux changing smoothly. As this scenario requires fine tuning, we consider it to be highly unlikely. 

We also point out that Xue et al.\ (2007) present another argument against the jet dominance in X-rays in the hard state. By modeling the spectrum of the jet blob, they obtain the parameters of the electron energy
distribution. Assuming the electron energy distribution is the same for
all blobs in the jet, they model the overall radio spectrum of the
source and extrapolate their fit to higher energies. They find that
the synchrotron emission from the jet contribute negligibly to the observed
X-ray emission. 

\acknowledgements
We thank John Tomsick and the referee, Chris Done, for valuable comments. 
This work was supported in part by the National Natural Science Foundation of China 
(grant 10543003) and One-Hundred-Talent Program of China (FY), 
and the Polish grants 1P03D01827, 1P03D01128 and 4T12E04727 (AAZ).

{}

\clearpage

\end{document}